\documentclass[aps,prb,twocolumn,superscriptaddress,amsfonts,amssymb,showpacs]{revtex4}

\usepackage{graphicx}
\usepackage{dcolumn}
\usepackage{amssymb}
\usepackage{wasysym}

\begin{document}


\title{NMR relaxation and rattling phonons in type-I Ba$_{8}$Ga$_{16}$Sn$_{30}$ clathrate}


\author{Xiang Zheng}
\affiliation{Department of Physics and Astronomy, Texas A\&M University, College
Station, TX 77843, USA}
\affiliation{Materials Science and Engineering Program,  Texas A\&M University, College Station, TX 77843, USA}
\author{Sergio Y. Rodriguez}
\affiliation{Department of Physics and Astronomy, Texas A\&M University, College Station, TX 77843, USA}
\author{Joseph H. Ross, Jr.}
\affiliation{Department of Physics and Astronomy, Texas A\&M University, College Station, TX 77843, USA}
\affiliation{Materials Science and Engineering Program,  Texas A\&M University, College Station, TX 77843, USA}


\date{June 7, 2011}

\begin{abstract}
Atomic motion of guest atoms inside semiconducting clathrate cages is considered as an important source for the glasslike thermal behavior.$^{69}$Ga and $^{71}$Ga  Nuclear Magnetic Resonance (NMR)  studies on type-I Ba$_{8}$Ga$_{16}$Sn$_{30}$ show a clear low temperature relaxation peak attributed to the influence of Ba rattling dynamics on the framework-atom resonance, with a quadrupolar relaxation mechanism as the leading contribution. The data are analyzed using a two-phonon Raman process, according to a recent theory involving localized anharmonic oscillators. Excellent agreement is obtained using this model, with the parameters corresponding to a uniform array of localized oscillators with very large anharmonicity. 
\end{abstract}

\maketitle

\section {INTRODUCTION}

Clathrates are materials with oversized polyhedral cages and guest atoms loosely bound inside. Type-I clathrates are based on the $A_8X_{46}$ structure, where $X$ stands for the cage atom and $A$ is the guest atom (Ba, Eu, Sr etc.). Group IV clathrates with cages formed by Ga, Ge, Sn and Si atoms are particularly well studied examples. Recent investigations of such intermetallic clathrates showed excellent thermoelectric properties\cite{1,2,12,SnyderNature2008,Dalton2010}. This is especially interesting for the group IV elements given their great potential in the  modern semiconductor industry \cite{3,4}. Research on the framework and caged atomic motions has also shown anomalous vibrational properties and their connection to the electronic behavior \cite{5,6}. Understanding the guest atom vibration modes, often called "rattling", has been considered to be one of the most important ways to reveal the essence of these phenomena. Many methods including Raman scattering \cite{7,TakasuPRB2010}, inelastic neutron scattering \cite{8}, optical conductivity \cite{9,10}, NMR relaxation \cite{11} and theoretical calculations \cite{dongJAP2000,GeorgPRB2005} have been reported to analyze the rattling atoms in clathrates using different models.

During the last few years, ultra-low lattice thermal conductivity ($\kappa_L$) and glasslike thermal behavior have been discovered in particular in type-I Ba$_{8}$Ga$_{16}$Sn$_{30}$ \cite{12,13,14}. The rattling of guest atoms in the larger of its two structural cages has been confirmed to have close connection to those properties \cite{13,14}. Anharmonic oscillators have also been used as trial models for the rattling phonons to analyze the thermoelectric properties and NMR relaxation behavior of other similar materials \cite{15,16,17}. In this paper, we discuss the NMR relaxation behavior and guest atomic motion of Ba$_{8}$Ga$_{16}$Sn$_{30}$ clathrates. A Raman process involving local vibrational modes, which is responsible for the NMR quadrupolar relaxation, will be discussed. Simple one and two dimensional anharmonic potentials will be introduced to model the anharmonic local oscillators. The shape and energy levels of these potential wells will be obtained by matching the simulations to the NMR experimental results.

\section{Sample preparation}

The clathrate samples were prepared using the self-flux method, following a technique reported previously \cite{18}. Because of the existence of a type-l/type-VIII dimorphism in Ba$_{8}$Ga$_{16}$Sn$_{30}$, a carefully controlled annealing process is needed during the sample making procedure. For type-I Ba$_{8}$Ga$_{16}$Sn$_{30}$, the pure elements were mixed together based on the nominal composition followed by an initial arc melting in argon environment. Annealing in an evacuated quartz tube at 900 $^{\rm o}$C for 50 hours was then applied, followed by a controlled slow cooling to 500 $^{\rm o}$C in 80 hours \cite{18,19}. X-ray diffraction (XRD) measurements were performed on a Bruker D8 X-ray Powder Diffractometer, and wavelength dispersion spectroscopy (WDS) measurements were done in a Cameca SX50 spectrometer.  Rietveld refinements of the XRD results were performed using EXPGUI \cite{24}, and the result confirmed the composition and structure of type-I Ba$_{8}$Ga$_{16}$Sn$_{30}$ with no type-VIII reflections detected and with 1\% (per mol) Ba(Ga,Sn)$_4$ minority phase obtained in the fit \cite{19}.

\begin{figure}
\includegraphics[width=\columnwidth]{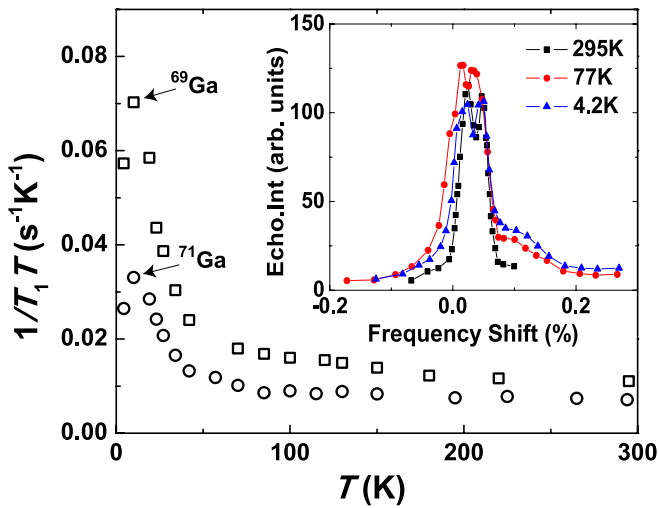}
\caption{\label{fig:fig1}  $^{71}$Ga and $^{69}$Ga NMR spin-lattice relaxation rates at the central transition frequency under 8.8 T from 4.2K to 295 K. Inset: $^{71}$Ga NMR lineshapes at temperatures 4.2 K, 77 K and 295 K, scaled proportional to 1/$T$.}
\end{figure}

\section{NMR results and discussion}
NMR experiments were carried out under external magnetic fields of 8.8 T and 7 T in a temperature range from 4.2 K to 295 K using a pulse spectrometer and a homemade multi-temperature detecting probe. The nuclei measured are $^{71}$Ga and $^{69}$Ga with different gyromagnetic ratios $\gamma$ and quadrupole moments Q, where $^{71}\gamma=8.1355$ rad/s G$^{-1}$,  $^{69}\gamma=6.4208$ rad/s G$^{-1}$ (ref. \onlinecite{27}),  $^{71}Q=10.7$ fm$^2$, and $^{69}Q=17.1$ fm$^2$ (ref. \onlinecite{26}).  The inset to Fig.1 shows the central portion of the  $^{71}$Ga NMR lineshapes at three temperatures under 8.8T. No significant change in average shift vs temperature has been observed in the lineshape mapping. The small changes at the base of the lineshape vs. temperature are due to unreacted Ga metal. The $^{71}$Ga lineshape is a superposition of two close peaks, which are due to different sites of the framework atoms \cite{19}. Here we consider only the behavior at the center of the resonance, which is due to a superposition of different local configurations. The weighted center shift of this resonance is about 0.033\% at 295 K. For comparison, the $^{69}$Ga lineshape under the same conditions, has a weighted shift of 0.023\%. The shift includes magnetic and quadrupole terms, which can be expressed as ${\delta}f=K+BQ^2$, where $K$ stands for the magnetic shift and $Q$ is the quadrupole moment of the nucleus. From the observed field-dependence we extracted $K=0.039\%$ as the center-of-mass magnetic shift and a negative quadrupole shift. In this case $K$ is mainly a Knight shift due to conduction electrons, with some contribution due to chemical shifts. Note that we did not observe any significant change in $K$, such as those observed in Na-Si type II clathrates \cite{GrykoPRB1998}, indicating excitations involving sharp electronic features in the system. As reported previously we have also performed a structural analysis using $ab$-$initio$ calculations of the low-temperature NMR shifts for this sample, modeling in particular the first-order quadrupole broadening at the base of the lineshape according to the distribution of Ga framework occupation. This agreed with the experimental lineshape quite well \cite{19}. 

NMR spin-lattice relaxation measurements were performed at the central transition frequency at the center of the lineshape. The relaxation time, $T_1$, is a fitted value based on a magnetic relaxation mechanism using a standard multi-exponential function for recovery of the central transition \cite{11}. The quadrupole relaxation process entails a different relaxation function, which however leads only to an overall scaling of the $T_1$, and does not affect any of the dynamical fitting parameters described below. The signal is also a superposition of different framework sites, however we fitted to a single average $T_1$ as parameter. Fig. 1 shows the resulting rates for both $^{71}$Ga and $^{69}$Ga under a field of 8.8 T. A clear peak at a temperature around 10 K can be observed for both nuclei.

The isotopic ratio, $^{69}T_{1}/^{71}T_{1}$, under 8.8 T is shown in the inset of Fig. 2. According to hyperfine relaxation theory \cite{21}, if the relaxation mechanism contains only a magnetic part, $T_1$ should be inversely proportional to ${\gamma}^2$ which gives  
$^{69}T_{1}/^{71}T_{1}{\cong}1.67$, while if the quadrupolar relaxation is in control, $T_1$ should be inversely proportional to 
$Q^2$ which gives  $^{69}T_{1}/^{71}T_{1}{\cong}0.4$. The experimental ratio for our sample is consistent over a wide temperature range and is  close to the quadrupole moment ratio. Thus, the relaxation is mainly controlled by the quadrupole mechanism, indicating that lattice vibrations are the most important contribution. As our experimental data are a mixture of magnetic and quadrupole parts, it is necessary to separate them for further investigations. The corresponding relationships are \cite{21},
\begin{eqnarray}
\frac{1}{T_1}=\frac{1}{T_{1Q}}+\frac{1}{T_{1M}}\\
\frac{1}{T_{1M}}\propto{\gamma}^2, \frac{1}{T_{1Q}}\propto\frac{1}{Q^2}
\end{eqnarray}
where $T_1$ is the overall experimental relaxation time while $T_{1M}$ and $T_{1Q}$ represent the magnetic and quadrupole parts. According to Eq. (1) and (2), the relaxation rates were separated into two contributions as shown in Fig. 2. Again, the result confirms the dominant role of the quadrupole relaxation rate. 

At higher temperatures, as values of $K^{2}T_{1}T$ do not change much, therefore the sample appears to follow a 
Korringa-like relation \cite{21,22}, which would normally indicate the influence of metallic electrons if $1/T_1$ 
were magnetic. However a recent model \cite{17} for relaxation dominated by anharmonic localized vibrations indicates such behavior as a high-temperature limit, along with a low-temperature peak much as observed here. Our later simulation based on this anharmonic model will be compared with the quadrupolar relaxation rates we have separated.

\begin{figure}
\includegraphics[width=\columnwidth]{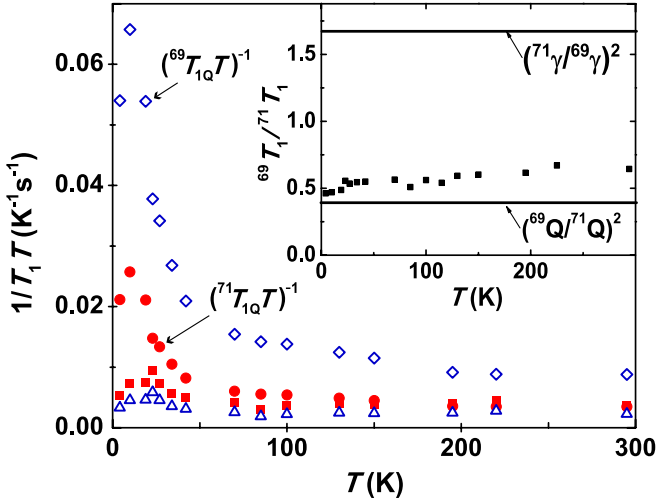}
\caption{\label{fig:fig2} Separated $T_1$ relaxation rates for $^{71}$Ga and $^{69}$Ga:  $^{69}$Ga-quadrupole (diamonds, as labeled), 
$^{71}$Ga-quadrupole (circles, as labeled), $^{71}$Ga-magnetic (squares), $^{69}$Ga-magnetic (triangles). 
Inset: isotopic ratio of overall  rates under 8.8 T, with limits for pure quadrupolar/magnetic relaxation indicated.}
\end{figure}

\section{Anharmonic model and fitting}
From refinements of x-ray diffraction spectra for Ba$_{8}$Ga$_{16}$Sn$_{30}$, the guest Ba(2) atom has location probability concentrated near four equivalent off-center positions with off-center dynamic displacements around 0.4 \AA \cite{13}. Our first principles calculations \cite{19} also gave similar values for the static displacement of the Ba atom due to cage asymmetry. A 1-D double well potential was also introduced by 
Dahm and Ueda to analyze this kind of problem, and has shown good agreement for the pyrochlore case \cite{17}. To model our data, therefore we use the Hamiltonian,
\begin{equation}
H=\frac{p^2}{2M}+\frac{1}{2}ax^2+\frac{1}{4}bx^4
\end{equation}
where $M$, $p$, and $x$ are the mass, momentum and spatial coordinate of the guest atom Ba \cite{15,17}. An effective localized phonon frequency, $\omega_0$, and thermal average of $x^2$ were introduced in a self-consistent quasiharmonic approximation giving $M{\omega}_0^2=a+b{\langle}x^2{\rangle}_{{\omega}_0,T}$, where
\begin{eqnarray}
{\langle}x^2{\rangle}_{{\omega}_0,T}=\frac{\hbar}{M{\omega}_0}\left(\frac{1}{e^{\hbar{\omega}_0/k_BT}-1}+\frac{1}{2}\right),
\end{eqnarray}
and the relationship between $\omega_0$ and $T$ is given by,
\begin{eqnarray}
\left(\frac{{\omega}_0}{{\omega}_{00}}\right)^2=1+\beta\frac{{\omega}_{00}}{{\omega}_0}\left(\frac{1}{e^{\hbar{\omega}_0/k_BT}-1}+\frac{1}{2}-\frac{{\omega}_0}{2{\omega}_{00}}\right),    
\end{eqnarray}
where ${\omega}_{00}={\omega}_0(T=0)$, and $\beta=b{\hbar}/M^2{\omega}_{00}^3$ is a dimensionless anharmonicity factor. 

As the relaxation is dominated by the quadrupole term, a two-phonon Raman process can be used to describe the NMR 
relaxation. This can be expressed \cite{17} as
\begin{eqnarray}
\frac{1}{T_1^R}=V_2^2{\int}_{-\infty}^{\infty}dt{\rm exp}\{i{\omega}_Lt\}{\langle}x^2(t)x^2(0)\rangle\nonumber\\=2\pi\left(\frac{\hbar}{2{\omega}_0M}\right)^2V_2^2{\int}_{-\infty}^{\infty}d{\omega}A^2(\omega)[n(\omega)+1]n(\omega)
\end{eqnarray}
with
\begin{eqnarray}
A(\omega)=-\frac{1}{\pi}{\rm Im}D(\omega)=\frac{1}{\pi}\frac{4{\omega}_0{\Gamma}_0\omega}{({\omega}^2-{\omega}_r^2)^2+4{\Gamma}_0^2{\omega}^2}
\end{eqnarray}
where $V_2$ is the second spatial derivative of the electric field gradient, ${\omega}_L$ is the nuclear 
Lamor frequency, $A(\omega)$ is the phonon spectral function, $n(\omega)$ is the Bose function, 
$D(\omega)$ is the retarded phonon propagator, 
$\Gamma_0$ is a phonon damping rate and ${\omega}_r^2$ is the renormalized phonon frequency 
determined by the phonon self-energy, ${\omega}_r^2={\omega}_0^2+2{\omega}_0$Re$\Pi(\omega)$. 
Here, we assume the real part of the phonon self energy, Re$\Pi(\omega)$, to be temperature independent 
as assumed in [17]. By carefully choosing parameters, we obtained a good fit to our data as shown in Fig. 3.  The matching results clearly show that the spin-lattice relaxation mechanism can be explained by the rattling phonon model.

\begin{figure}
\includegraphics[width=\columnwidth]{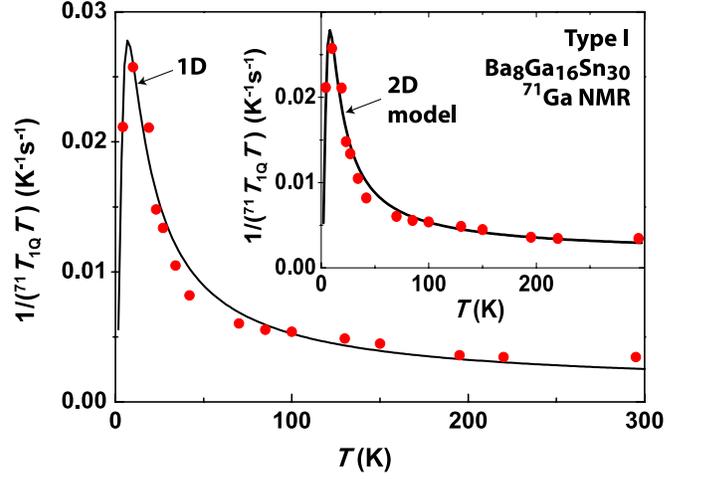}
\caption{\label{fig:fig3} Quadrupole NMR relaxation rate for $^{71}$Ga compared with the fitted 1-D 
anharmonic model (main plot, solid curve) and simplified 2-D model (inset, solid curve).}
\end{figure}

The corresponding values for the parameters are ${\omega}_{00}=20$ K, $\beta=50$, 
${\Gamma}_0=12$ K and ${\omega}_r(T=0)=19.5$ K. The potential well is given by the calculated expression,
\begin{eqnarray}
V(x)=-18.74 x^2+1.11\times10^{23} x^4, 
\end{eqnarray}
where $V(x)$ is in J with $x$ given in meters. Also from equations (4) and (5), when $T=296$ K, 
${\omega}_0$ ${\cong}$ 11 THz and ${\langle}x^2{\rangle}_{{\omega}_0,T}^{1/2}$ ${\cong}$ 0.12 {\AA}. Since in 2D 
${\langle}r^2{\rangle}_{{\omega}_0,T}=2{\langle}x^2{\rangle}_{{\omega}_0,T}$, this corresponds to a 
rms guest atom displacement of 0.17 {\AA}, which is not far from the values reported previously \cite{13,19}. 
Solving the Schr{\"o}dinger equation numerically, the energy levels of this double well potential can 
be calculated as shown in Fig. 4. The energy difference between the lowest two states, 
${\Delta}E_{12}$ ${\cong}$ 30 K, is much smaller than those for higher energy levels, which 
agrees with recent reported results from other methods \cite{9}, but with a larger ${\Delta}E_{12}$. 

\begin{figure}
\includegraphics[width=\columnwidth]{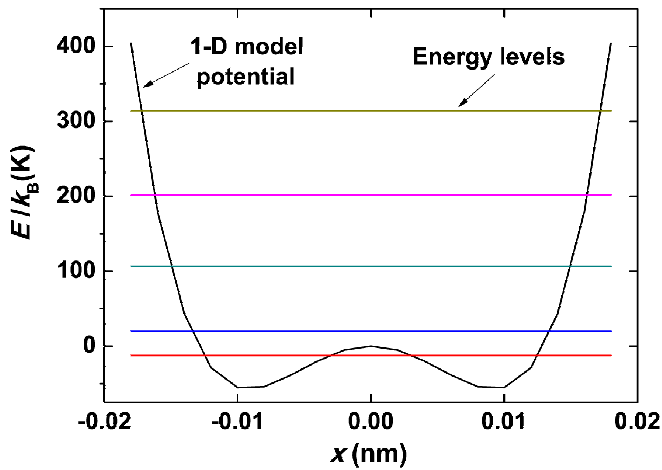}
\caption{\label{fig:fig4} Fitted 1-D double well potential and its energy levels.}
\end{figure}

This model can also be extended to a simplified 2-D potential by using 
${\langle}r^2{\rangle}_{{\omega}_0,T}={\langle}x^2{\rangle}_{{\omega}_0,T}+{\langle}y^2{\rangle}_{{\omega}_0,T}=
2{\langle}x^2{\rangle}_{{\omega}_0,T}$ in equation (4). We correspondingly modified 
the relationship in equation (5). Then, following the same procedure, we obtain a 
fitting similar to the 1-D model. The result is shown in the inset of Fig. 3, with the fitted values 
${\omega}_{00}=20$ K, $\beta=25$, ${\Gamma}_0=12$ K and ${\omega}_r(T=0)=19.5$ K. 
The corresponding potential is $V(r)=-8.98r^2+5.52\times10^{22}r^4$, where $V(r)$ is in 
J with $r$ given in meters. The average displacement is still 0.17 \AA, which indicates 
a consistency of the model compared with the 1-D case.

Compared to previous Ga NMR results for Sr$_{8}$Ga$_{16}$Ge$_{30}$, also identified to behave as an anharmonic rattler system \cite{2,dongJAP2000}, it seems initially surprising that the $(T_1T)^{-1}$ in Sr$_{8}$Ga$_{16}$Ge$_{30}$ 
does not show a similar phonon-dominated behavior but instead follows a Korringa law 
quite closely for several decades of temperature \cite{11}. However, a previous report for 
Sr-Ge clathrates used density functional theory to extract potential well parameters for Sr 
in the large cage \cite{dongJAP2000}, giving a 2D anharmonic potential much like the 
model used here. The resulting potential has a very small quadratic term, but also a much smaller 
anharmonicity parameter $(\beta)$ than found here, and inserting the calculated parameters into 
the relaxation theory described above yields a smaller quadrupole contribution to $(T_1T)^{-1}$, 
rising slowly with temperature without exhibiting a peak as in Fig. 3. Also in the analysis reported 
for the elastic response of Eu- and Sr-filled Ge clathrates \cite{23} a four-well potential was used 
to model the vibrational response for Eu, but for Sr a harmonic Einstein oscillator model provided 
satisfactory agreement. Thus it is consistent that the $(T_1T)^{-1}$ in Sr$_{8}$Ga$_{16}$Ge$_{30}$ 
is dominated by interactions with the charge carriers, while the quadrupole-dominated peak observed 
in Ba$_8$Ga$_{16}$Sn$_{30}$ is indicative of the much larger anharmonicity for rattler atoms. 

In our measurements, we prepared a second sample of type-I Ba$_8$Ga$_{16}$Sn$_{30}$ in the 
same way, whose $^{71}T_1^{-1}$ exhibits a low-temperature maximum that is nearly identical to 
that of Fig. 1. In fitting to the model oscillator potential, the position of the $^{71}T_1^{-1}$ peak is
particularly sensitive to ${\omega}_{00}$, which is close to the spacing of the two lowest levels in 
Fig. 4. The ability to model this behavior with a single set of parameters attests to the lack of irregularity 
of the cage potential, despite the presence of quasi-random framework substitution. This is 
apparently due to the Sn-based cage size, providing space for the relatively unconstrained 
motion of the Ba(2) atoms without allowing for a permanent distortion \cite{13}. Thus, type-I 
Ba$_8$Ga$_{16}$Sn$_{30}$ can be viewed as possessing a more or less uniform array of 
strongly anharmonic local oscillators. The NMR relaxation times are particularly sensitive to 
the low-frequency anharmonic motion of these atoms, and thus provide an excellent probe for this behavior.

\section{Conclusion}

In conclusion, type-I Ba$_8$Ga$_{16}$Sn$_{30}$ NMR lineshapes and spin-lattice 
relaxation rates indicated the presence of a strong quadrupole relaxation mechanism. 
Analysis showed this behavior to be due to a strongly anharmonic rattler-type motion of 
the caged Ba atoms. Fitting using a 1-D double well potential with strong anharmonicity 
showed good agreement with the experimental data, which offers a good explanation 
for the rattling behavior and the relaxation mechanism.

\begin{acknowledgments}
This work was supported by the Robert A. Welch Foundation, Grant
No. A-1526, by the National Science Foundation (DMR-0103455).
\end{acknowledgments}

\end{document}